\documentclass[twocolumn,showpacs,preprintnumbers,amsmath,amssymb]{revtex4}
\usepackage{graphicx}
\usepackage{epsfig}  
\usepackage{epsf}    
\usepackage{dcolumn}
\usepackage{bm}

\def\issue(#1,#2,#3){{\bf #1}, #2 (#3)} 

\def\opcit(#1){ {\em op. cit.}, #1}
\def\etal {\em et al.}

\def\APP(#1,#2,#3){Acta Phys.\ Polon.\ \issue(#1,#2,#3)}
\def\ARNPS(#1,#2,#3){Ann.\ Rev.\ Nucl.\ Part.\ Sci.\ \issue(#1,#2,#3)}
\def\CPC(#1,#2,#3){Comp.\ Phys.\ Comm.\ \issue(#1,#2,#3)}
\def\CIP(#1,#2,#3){Comput.\ Phys.\ \issue(#1,#2,#3)}
\def\EPJC(#1,#2,#3){Eur.\ Phys.\ J.\ C\ \issue(#1,#2,#3)}
\def\EPJD(#1,#2,#3){Eur.\ Phys.\ J. Direct\ C\ \issue(#1,#2,#3)}
\def\IEEETNS(#1,#2,#3){IEEE Trans.\ Nucl.\ Sci.\ \issue(#1,#2,#3)}
\def\IJMP(#1,#2,#3){Int.\ J.\ Mod.\ Phys. \issue(#1,#2,#3)}
\def\JHEP(#1,#2,#3){J.\ High Energy Physics \issue(#1,#2,#3)}
\def\MPL(#1,#2,#3){Mod.\ Phys.\ Lett.\ \issue(#1,#2,#3)}
\def\NP(#1,#2,#3){Nucl.\ Phys.\ \issue(#1,#2,#3)}
\def\NIM(#1,#2,#3){Nucl.\ Instrum.\ Meth.\ \issue(#1,#2,#3)}
\def\PL(#1,#2,#3){Phys.\ Lett.\ \issue(#1,#2,#3)}
\def\PRD(#1,#2,#3){Phys.\ Rev.\ D \issue(#1,#2,#3)}
\def\PRL(#1,#2,#3){Phys.\ Rev.\ Lett.\ \issue(#1,#2,#3)}
\def\SJNP(#1,#2,#3){Sov.\ J. Nucl.\ Phys.\ \issue(#1,#2,#3)}
\def\ZPC(#1,#2,#3){Zeit.\ Phys.\ C \issue(#1,#2,#3)}

\def\bra {\langle}
\def\ket {\rangle}

\def\l {\lambda}

\def\r {\rightarrow}

\def\bar {\overline}
\def\bbbar {B^0-\bar{B^0}}
\def\kkbar {K^0-\bar{K^0}}
\def\mmbar {M^0-\bar{M^0}}
\def\bsbsbar {B_s-\bar{B_s}}
\def\msbar {\overline{MS}}

\def\be {\begin{equation}}
\def\ee {\end{equation}}
\def\bea {\begin{eqnarray}}
\def\eea {\end{eqnarray}}

\def\bc {\begin{center}}
\def\ec {\end{center}}

\begin{document}

\preprint{CU-PHYSICS-02/2004}
\preprint{hep-ph/0403154}

\title{Constraints on R-parity violating supersymmetry from neutral
meson mixing}

\author{Anirban Kundu}
\affiliation{Department of Physics, University of Calcutta,\\
92 A.P.C. Road, Kolkata 700009, India}
\email{akundu@cucc.ernet.in}

\author{Jyoti Prasad Saha}
\affiliation{Department of Physics, Jadavpur University, Kolkata 700032, India}
\email{jyotip@juphys.ernet.in}

\date{\today}

\begin{abstract}

Upper bounds at the weak scale are put on all $\lambda'_{ijk}\lambda'_{imn}$ 
type products of R-parity violating supersymmetry that may affect
$\kkbar$ and $\bbbar$
mixing. We constrain all possible products, including some not considered 
before, using next-to-leading order QCD corrections to the mixing amplitudes.
Constraints are obtained for both real and imaginary parts of the couplings.
We also discuss briefly some correlated decay channels which should be
investigated in future experiments.

\end{abstract}


\maketitle

\section{Introduction}

Is there any new physics (NP) beyond the Standard Model (SM)? Probably
yes, if high-energy physics does not want itself to be lost in a mire
of unanswered questions. Supersymmetry (SUSY) is one of the most widely 
discussed options of NP, in both its R-parity conserving (RPC) and 
R-parity violating (RPV) incarnations.  However, SUSY introduces
a plethora of new particles, and even in its most constrained version, a few
more arbitrary input parameters over and above to that of the SM. Thus, it
has become imperative to constrain the SUSY parameter space as far as possible
from existing data. 

There are, of course, direct bounds on the sparticle masses from collider 
data. These bounds are typically weaker for RPV SUSY than for the RPC
version, since in the former case the final-state signal is radically different
and more possible channels are open. Apart from them, there are bounds on
the parameter space from low-energy processes, whose amplitudes can be 
affected by intermediate sparticle states. For RPC SUSY, the new amplitudes
must be at least at one-loop level, so that they can compete with the SM 
amplitude and show up only if the SM amplitude is also suppressed. The new
amplitudes may appear at tree-level if R-parity is violated, and hence there
is a greater chance of constraining RPV SUSY models from low-energy data.
There is another good reason to focus upon the RPV version: most of the
precision low-energy processes from which one can obtain bounds are
flavor-changing neutral current (FCNC) type. Now, there are some elegant
mechanisms of suppressing FCNC effects in RPC SUSY, but absolutely none (except
putting some flavor-changing Yukawa-type couplings equal to zero, or 
vanishingly small, by hand) if we have RPV. Thus, the bounds on the parameter
space of RPV SUSY coming from FCNC processes are in a sense more robust. 

In this paper we use the data from $\kkbar$ and $\bbbar$ mixing to constrain
the relevant couplings for RPV SUSY. For the latter we use both $\Delta m_B$
and $\sin(2\beta)$ constraints, while for the former we use the
results on $\Delta m_K$ and $\varepsilon_K$. We do not discuss other CP
violating parameters like $\varepsilon'/\varepsilon$, since that has large
theoretical uncertainty.  We have also
discussed some correlated channels leading to
leptonic and semileptonic decays.  We do not consider the
$B_s$ system since there is only a lower bound on $\Delta M_{B_s}$. This,
in turn, means that there is no such upper bound on the relevant NP couplings;
we can only have a lower bound, which is consistent with zero. The situation
should change dramatically once the hadronic B machines, producing copious
$B_s$ mesons, come on line.

Do we have any motivation to invoke NP for the K and the B systems? In other
words, is there any inconsistency of the experimental data with the SM
predictions? The answer is yes for the B system, though the error bars are still
large to draw any definite conclusion (but we have reasons to be hopeful).
The sore thumbs are  
(i) the abnormally high branching ratios (BR) for the
generic channels $B\r \eta' K, \eta K^*$ \cite{hfag} \footnote{We use $B^0$
and $\overline{B^0}$ for flavor eigenstates, $B_d$ for a state whose
flavor composition is immaterial, and B for a generic $B^0/\overline{B^0}/
B^\pm$ meson.}, (ii) the direct
CP-asymmetry in the channel $B_d\r \pi^+\pi^-$ as found by Belle \cite{pipi},
(iii) the discrepancy in the extracted value of $\sin(2\beta)$ from
$B_d\r J/\psi K_S$ and $B_d\r \phi K_S$ \cite{phiks}, and (iv) the so-called
``$\pi K$'' puzzle: the abnormal enhancement of electroweak penguins in
$B\to \pi K$ decays \cite{hfag} \footnote{This may not be a puzzle. See,
{\em e.g.}, S. Nandi and A. Kundu, hep-ph/0407061.}. However, one must not
be over-enthusiastic since these channels are nonleptonic and QCD uncertainties
are yet to be fully understood. But one may hope more such anomalies from
leptonic and hadronic B-factories. For the K system, the nonleptonic channels
are notoriously difficult for any systematic analysis of NP effects
\cite{masiero:eps}, but 
for the first time we are having precise data (or bound) on leptonic and 
semileptonic K decay channels from Brookhaven and Da$\Phi$ne. It is
always better to be ready for any unexpected result.

Another motivation for supersymmetry comes from the neutrino mass. It has
been shown that the existence of nonzero neutrino mass may have observable
signatures in B-physics \cite{chang,nierste} like the enhancement of the
$b\to s\bar{s}s$ amplitude and also $\Delta M_s$, the mass difference between
two $B_s$ mass eigenstates, if one assumes some unified RPC SUSY theory. In
RPV SUSY, one may have a definite texture at the GUT scale (only a few RPV
couplings are nonzero) and can generate the whole lot of nonzero RPV couplings
at the weak scale through renormalization group evolution and CKM-type
mixing \cite{dreiner}. One
expects a signature of those RPV couplings relevant for neutrino mass 
generation in the B-factory data. 

Must the RPV couplings be complex? The answer is that there is no a priori
reason why they should be all real. Even if there is some real GUT texture,
the weak scale couplings may turn out to be complex. The phase of one single 
coupling can be absorbed in the sfermion field, but a nontrivial phase
should be there in a product of two such couplings. 

Effects of RPV SUSY on K and B physics have been discussed extensively in
the literature \cite{rpv:k,bsb,agashe,abel,rpv:b}. 
Constraints coming from $\kkbar$ and 
$\bbbar$ mixing have been discussed in \cite{gg-arc,decarlos-white,jyoti2}. This
paper is a culmination of the effort in the following sense: (i) This is 
the first time that both SM and all RPV contributions (including their 
phases) have been taken
into account for the systems under consideration; (ii) the short-distance
QCD corrections have been implemented upto next-to-leading order (NLO)
accuracy (there is scope for questioning this procedure for the K system; this
will be discussed later); (iii) Both CP-conserving
and CP-violating constraints have been discussed for the $\bbbar$ and
$\kkbar$ systems;
(iv) Some discrepancies in the relevant formulae used in some of the
earlier works have been corrected; and (v) Correlated signals for leptonic
and semileptonic decays have been discussed. 

The paper is arranged as follows. In Section 2 we outline the relevant formulae
necessary for the analysis. Section 3 deals with the numerical inputs and 
Section 4 with the analysis of the results. Decay channels mediated by
the same couplings are touched upon in Section 5,
while we conclude and summarize in Section 6. Some 
calculational details have been relegated to the two appendices.

\section{Basic inputs}
\subsection{Neutral meson mixing}

Let the neutral mesons be generically denoted by $M^0$ and $\bar{M^0}$, with
the valence quark content $\bar{q} d$ and $q\bar{d}$ respectively. For the
cases under study, $q$ can be either $d$ or $s$. For $B_s$ system, replace
$q$ by $b$ and $d$ by $s$.  

The off-diagonal element in the $2\times 2$ effective Hamiltonian 
causes the $\mmbar$ mixing.  The mass difference
between the two mass eigenstates $\Delta M$ is given by (following
the convention of \cite{buras-fleischer})
\begin{equation}
\Delta M = 2|M_{12}|,
\end{equation}
with the approximation $|M_{12}|\gg |\Gamma_{12}|$. This, however, is true for
the B system only.
Let the SM amplitude be
\begin{equation}
|M_{12}^{SM}|\exp(-2i\theta_{SM})
\end{equation}
where $\theta_{SM}= \beta (\phi_1)$ for the ${\bbbar}$ system and 
approximately zero
for the $\kkbar$ (and also for $\bsbsbar$) system. We follow the $(\alpha,
\beta,\gamma)$ convention for the unitarity triangle \cite{buras-fleischer}.

If we have $n$ number of NP amplitudes with weak phases $\theta_n$, one can
write
\begin{equation}
M_{12} = |M_{12}^{SM}|\exp(-2i\theta_{SM}) + \sum_{i=1}^n
|M_{12}^i|\exp(-2i\theta_i).
\end{equation}
This immediately gives the effective mixing phase $\theta_{eff}$ as
\begin{equation}
\theta_{eff} = {1\over 2}\arctan {|M_{12}^{SM}|\sin(2\theta_{SM}) + 
\sum_i|M_{12}^i|\sin(2\theta_i)
\over |M_{12}^{SM}|\cos(2\theta_{SM}) + \sum_i|M_{12}^i|\cos(2\theta_i)},
\end{equation}
and the mass difference between mass eigenstates as
\begin{eqnarray}
\Delta M & = & 2\Large[ |M_{12}^{SM}|^2 + \sum_i|M_{12}^i|^2\nonumber\\
& +& 2|M_{12}^{SM}|\sum_i |M_{12}^i|\cos 2(\theta_{SM}-\theta_i)\nonumber \\
& + & 2 \sum_i \sum_{j>i} |M_{12}^j||M_{12}^i|\cos 2(\theta_j-\theta_i)
\Large]^{1/2}.
\end{eqnarray}
These are going to be our basic formulae. The only task is to find
$M_{12}^i$ and $\theta_i$.

For the $\kkbar$ system \cite{paschos}, $|\Gamma_{12}|$ is non-negligible, and
we can write
\begin{eqnarray}
\Delta M&=&2~Re\left[ (M_{12}-{i\over 2}\Gamma_{12}) 
(M^*_{12}-{i\over 2}\Gamma^*_{12})\right]^{1/2},\nonumber\\
\Delta \Gamma&=&-4~Im\left[ (M_{12}-{i\over 2}\Gamma_{12}) 
(M^*_{12}-{i\over 2}\Gamma^*_{12})\right]^{1/2},
\end{eqnarray}
so that $\Delta M = -(1/2)\Delta \Gamma$. Since the dominant decay is to the
$I=0$ final state, one can neglect $Im\Gamma_{12}$ and write
\begin{equation}
\Delta M = 2~Re~M_{12}, \ \ \Delta \Gamma=2~Re~\Gamma_{12}.
\end{equation}
The CP-violating parameter $\varepsilon_K$ is given by
\begin{equation}
|\varepsilon_K| = {1\over 2\sqrt{2}} {Im~M_{12}\over Re~M_{12}},
\end{equation}
which can be written as
\begin{equation}
|\varepsilon_K| = {1\over \sqrt{2}} {Im~M_{12}\over \Delta M}.
    \label{epsa-k}
\end{equation}
Note that $Re~M_{12}$ has both short-distance (SD) and long-distance (LD)
contributions. The LD contribution is not calculable; what one calculates
from the box amplitude is the SD part. That is why one generally uses the
experimental value of $\Delta M_K$ in the denominator of eq. (\ref{epsa-k}).

For $\kkbar$ system, the short-distance SM amplitude is
\begin{eqnarray}
M_{12}^{SM}&\equiv& {\bra \bar{K^0}|H_{eff}|K^0\ket\over  2m_K}\nonumber\\
&\approx& {G_F^2\over 6\pi^2}(V_{cd}V_{cs}^*)^2
\eta_K m_K f_K^2 B_K m_W^2 S_0(x_c),
    \label{k-sm}
\end{eqnarray}
where generically $x_j = m_j^2/m_W^2$, $f_K$ is the K meson decay constant, 
and $\eta_K$ (also called $\eta_{cc}$ in the literature)
and $B_K$ parametrize the short- and the long-distance QCD
corrections, respectively ($B_K$ is a phenomenological parameter, thrown in
to parametrize the LD contribution). 
The tiny top-quark loop dependent part responsible
for CP violation has been neglected. The function $S_0$ is given by
\begin{equation}
S_0(x) = {4x-11x^2+x^3\over 4(1-x)^2} - {3x^3\ln x \over 2(1-x)^3}.
\end{equation}
For the $\bbbar$ system, we have an analogous equation, dominated by the
top quark loop:
\begin{eqnarray}
M_{12}^{SM}&\equiv& {\bra \bar{B^0}|H_{eff}|B^0\ket\over  2m_B}\nonumber\\
&=& {G_F^2\over 6 \pi^2}(V_{td}V_{tb}^*)^2
\eta_B m_B f_B^2 B_B m_W^2 S_0(x_t).
    \label{b-sm}
\end{eqnarray}
There is not enough motivation to consider the $\bsbsbar$ system right now,
since there exists only a lower bound on $\Delta M_s (\geq 14.4 
{\rm ps}^{-1})$ \cite{hfag}. This can accomodate
arbitrarily large NP couplings, starting from zero. However, if the NP
amplitude has a nonzero phase, then there will be an effective phase in 
$\bsbsbar$ mixing amplitude, whose presence may be tested in the hadronic 
B factories. Whether there is any detectable new physics in $\bsbsbar$ 
mixing (this is
particularly relevant since the $b\to s$ penguin transition, at least for
nonleptonic decays, shows hint of an anomalous behaviour) can be effectively 
tested in hadronic B machines, which will measure $\Delta M_s$ as well as
the CP asymmetries in $B_s\to J/\psi \phi$ and hopefully $B_s\to \phi
(\eta')\phi(\eta')$. Detection of NP signals in the double vector meson 
modes requires angular analysis of the decay products but hopefully can be
done in future colliders \cite{sinha}.

In the presence of NP, the
general $\Delta F = 2$ effective Hamiltonian can be written as
\begin{equation}
{\cal H}_{eff}^{\Delta F = 2} = \sum_{i=1}^5 c_i(\mu) O_i(\mu) + \sum_{i=1}^3
\tilde c_i(\mu) \tilde O_i(\mu) + H.c.
\end{equation}
where $\mu$ is the regularization scale, and
\begin{eqnarray}
O_1&=&(\bar q \gamma^\mu P_L d)_1(\bar q \gamma_\mu P_L d)_1,\nonumber\\
O_2&=&(\bar q P_R d)_1(\bar q  P_R d)_1,\nonumber\\
O_3&=&(\bar q P_R d)_8(\bar q  P_R d)_8,\nonumber\\
O_4&=&(\bar q P_L d)_1(\bar q  P_R d)_1,\nonumber\\
O_5&=&(\bar q P_L d)_8(\bar q  P_R d)_8,
   \label{operators}
\end{eqnarray}
where $q$ is either $b$ or $s$,
and $P_{R(L)}=(1+(-)\gamma_5)/2$. The subscripts 1 and 8 indicate whether
the currents are in color-singlet or in color-octet combination. The $\tilde
O_i$s are obtained from corresponding $O_i$s by replacing $L\leftrightarrow R$.
The Wilson coefficients $c_i$ at $q^2=m_W^2$ include NP effects, coming from
couplings and internal propagators. However, for most of the NP models,
and certainly for the case we are discussing here, all NP particles are
heavier than $m_W$ and hence the running of the coefficients between $m_W$
and $\mu={\cal O}(m_s~{\rm or}~m_b)$ are controlled by the SM Hamiltonian alone.
In other words, NP determines only the boundary conditions of the 
renormalization group (RG) equations.  
For the evolution of these coefficients down to the low-energy scale, we
follow Ref.\ \cite{becirevic}, which uses, for $\bbbar$ mixing, $\mu = m_b 
= 4.6$ GeV. 
The low-scale Wilson coefficients, using the NLO-QCD corrections, are
\begin{equation}
c_i(\mu) = \sum_r \sum_s \left(b_r^{i,s} + \eta d_r^{i,s}\right)\eta^{a_r}
c_s(M_S)
     \label{running-b}
\end{equation}
where $\eta = \alpha_s(M_S)/\alpha_s(m_t)$, $M_S$ being the scale of
NP, which, for SUSY, may be taken to be the average of the squark and the 
gluino masses. We will use $M_S = 500$ GeV throughout the paper. 
For the numerical values of
$a$, $b$ and $d$ matrices we refer the reader to eq.\ (10) of Ref.\
\cite{becirevic} (to avoid confusion with the Wilson coefficients, we use the
symbol $d$ for the matrix denoted by $c$ in \cite{becirevic}). 
All the numbers are not relevant for our discussion; 
the modification of the NP operators due to the short-distance QCD 
corrections is discussed in Section 4. It will be shown that only operators
$\tilde{O_1}$ and $O_4$ are relevant at the scale $q^2=m_W^2$, while 
there is a slight admixture of $O_5$ at the low-energy scale.

The operators $O_i$ are also to be renormalized at the scale $\mu$.
The expectation values of these operators between $\bar{B^0}$ and $B^0$
at the scale $\mu$ are given by
\begin{eqnarray}
\bra O_1(\mu)\ket &=& {2\over 3} m_B^2 f_B^2 B_1(\mu),\nonumber\\
\bra O_2(\mu)\ket &=& -{5\over 12} S_B m_B^2 f_B^2 B_2(\mu),\nonumber\\
\bra O_3(\mu)\ket &=& {1\over 12} S_B m_B^2 f_B^2 B_3(\mu),\nonumber\\
\bra O_4(\mu)\ket &=& {1\over 2} S_B m_B^2 f_B^2 B_4(\mu),\nonumber\\
\bra O_5(\mu)\ket &=& {1\over 6} S_B m_B^2 f_B^2 B_5(\mu),
    \label{bparam}
\end{eqnarray}
where
\begin{equation}
S_B = \left({m_B\over m_b(m_b) + m_d(m_b)}\right)^2
\end{equation}
The $B$-parameters, whose numerical values are given in Section 3, have been
taken from \cite{lattice-b}. Note that the expectation values are scaled by
factor of $2m_B$ over those given in some literature due to our different
normalization of the meson wavefunctions. It is trivial to check that
both conventions yield the same values for physical observables. Analogous
expressions follow for the $\kkbar$ system, with $m_B,f_B\to m_K,f_K$, and
\begin{equation}
S_K = \left({m_K\over m_s(\mu) + m_d(\mu)}\right)^2,
\end{equation}
with $\mu = 2$ GeV.
The running of the Wilson coefficients is given by an equation which is
exactly analogous to eq.\ (\ref{running-b}) \cite{ciuchini} but with
different $a$, $b$ and $d$ matrices, which are evaluated at $\mu=2$ GeV.
The $B$-parameters, which include all nonperturbative effects below
the scale $\mu$, are different too.  

It is noteworthy that the NLO corrections as discussed above
cannot be applied when there
are light quarks like $u$ flowing in the loop, since the operator
product expansion takes a different form \cite{ciuchini}. This will be
the case when we switch on the RPV interaction. There should be an
enhancement coming from large logarithms; however, this puts on a tighter
constraint on the RPV couplings. Since these QCD corrections are not
known, we use the same procedure as adopted for heavy particles, and
probably deduce somewhat lenient constraints on the parameter space for
those particular couplings.

It is clear from equations (\ref{k-sm}) and (\ref{b-sm}) that at least in
the SM, one can either evaluate the matrix element of $O_1$ at $m_W$
and compute the running to the low energy scale, or alternatively can
use the parameters $B_{K,B}$ and $\eta_{K,B}$, to be determined from the
lattice studies. We adopt the latter method for calculating the SM
amplitudes, scanning over the whole range of these parameters, so that
even the nonperturbative effects can be successfully taken into account,
including all the uncertainties. This gives us the most conservative
bounds on the NP parameters. The NP amplitude is calculated using the
former method.

\subsection{R-Parity violating SUSY}

R-parity is a global quantum number, defined as $(-1)^{3B+L+2S}$, which
is $+1$ for all particles and $-1$ for all superparticles. In the minimal
version of supersymmetry and some of its variants, R-parity is assumed
to be conserved {\em ad hoc}, which prevents single creation or
annihilation of superparticles. Such models have a natural candidate for
dark matter, and have telltale signatures in collider searches as large
missing energy. However, models with broken R-parity can
be constructed naturally, and such models have a number of interesting
phenomenological consequences \cite{rpvrefs1,rpvrefs2}.
Some of these R-parity violating models can be motivated from an underlying
GUT framework \cite{rpvgut}.

It is well known that in order to avoid rapid proton decay one cannot
have both  lepton number and  baryon number violating RPV couplings, and we
shall work with a lepton number violating model. This leads
to both slepton (charged and neutral) and squark  
mediated decays, and new amplitudes for $\mmbar$ mixing, where particles 
flowing inside the box can be (i) charged sleptons and up-type
quarks, (ii) sneutrino and down type quarks, (iii) squarks and leptons. One
or both of the scalar particles inside the box can be replaced by 
$W$ bosons, charged Higgs bosons and Goldstone bosons (in a non-unitary gauge) 
(see Fig.\ 1). We follow the usual practice of avoiding the so-called ``pure
SUSY'' contributions to the box amplitudes, {\em i.e.}, those coming from
charginos, neutralinos or gluinos inside the loop. Not only the strongly
interacting superparticles are expected to be heavier than the electroweak
ones (and hence the contribution being suppressed), but also one can choose
SUSY models where these contributions become negligible ({\em e.g.}, alignment
in the squark sector, or Higgsino-dominated
lighter chargino, to kill off the respective boxes.) 
Since the current lower bound on the slepton mass is generally
weaker than that on squark mass by a factor 2-3, the slepton mediated
boxes have greater chance to be numerically significant.

We start with the superpotential
\begin{equation}
\label{w}
{\cal W}_{\lambda'} = \lambda'_{ijk} L_i Q_j D^c_k,
\end{equation}
where $i,j,k = 1,2,3$ are quark and lepton generation indices;
$L$ and $Q$ are the $SU(2)$-doublet lepton and quark superfields and
$D^c$ is the $SU(2)$-singlet down-type quark
superfield respectively. Written in terms of component fields, this
superpotential generates six terms, plus their hermitian conjugates:
\begin{widetext}
\begin{equation}
{\cal L}_{LQD} = \l'_{ijk} \left[ \tilde\nu^i_L {\bar d}^k_R d^j_L
+ \tilde d^j_L {\bar d}^k_R \nu^i_L + (\tilde d^k_R)^* {\bar\nu}^i_L
d^j_L
- \tilde e^i_L {\bar d}^k_R u^j_L - \tilde u^j_L {\bar d}^k_R e^i_L
-(\tilde d^k_R)^* {\bar e }^i_L u^j_L\right] + H.c.
\end{equation}
\end{widetext}
With such a term, one can have two different kind of boxes, shown in Fig.\ 1,
that contribute
to $\mmbar$ mixing: first, the one where one has two sfermions flowing inside
the loop, alongwith two SM fermions \cite{decarlos-white}, and secondly,
the one where one slepton, one $W$ (or charged Higgs or Goldstone) and two
up-type quarks complete the loop \cite{gg-arc}. It is obvious that the first
amplitude is proportional to the product of four $\lambda'$ type
couplings, and the second to the product of two $\lambda'$ type couplings
times $G_F$. We call them L4 and L2 boxes, respectively, for brevity,
where L is a shorthand for $\l'$.

We will constrain only products of two $\lambda'$-type couplings at a time,
and assume a hierarchical structure, {\em i.e.}, only one product is,
for all practical purpose, simultaneously nonzero (but can have a nontrivial 
phase). This may not be physically 
the most appealing scenario but keeps the discussion free from unnecessary
complications. For any product, there are individual bounds on each of them,
the product of which we call the direct product bound (DPB). Interesting
bounds are those which are numerically smaller, and hence stronger,
than the corresponding DPBs, the more the
better. This turns out to be the case for almost all the products. A list
of the individual bounds at the weak scale can be found in \cite{dreiner}.

At this point, let us clarify how the bounds are obtained. There are, 
generically, three boxes: SM, L2 and L4. 
For L2 boxes, one can have same or different up-type quarks flowing inside
the loop. For the former case, there is also an L4 counterpart, {\em i.e.},
the coupling that gives rise to the L2 box can also generate an L4 box.
The operator for the L2 box is $O_4$, while that for the L4 box is 
$\tilde{O_1}$. The RG evolution generates a small $O_5$ admixture to $O_4$
at the low-energy scale, while $\tilde{O_1}$ is just multiplicatively
renormalized. 

The effective Hamiltonian is obtained from a coherent sum of these three
amplitudes, and there can be intricate interference patterns
depending on the magnitudes and the phases of them. A general
trend, however, is easy to follow. 
The pure SM part is
   proportional to $G_F^2$ times some CKM factor. The L2 box is proportional
   to $G_F{\l'}^2/{\tilde m}^2$, and the
   L4 box goes as ${\l'}^4/{\tilde m}^4$. 
For small values of the $\lambda'\lambda'$ 
coupling, L2 contribution dominates over L4, and the ``bound'' (as
obtained by \cite{gg-arc})  is controlled by
L2. We will see that if the phases are included, this procedure does not
give the bounds, so the numbers obtained in this way are not listed in
tables 1 and 2. 
For larger values of the coupling, L4 becomes dominant, and if there 
happens to be a cancellation between SM and L2 amplitudes, L4 controls the
show. That is, the solution describes a contour in the
   $Re(\lambda'\lambda')-Im(\lambda'\lambda')$ plane. There are certain points,
far from the origin, where there happens to be a complete destructive 
interference between SM and L2 boxes.
Thus, if the product has an arbitray phase, it is always the L4 term
that gives the bound when the quarks inside are the same. 
We will see explicit examples of this later.  

If the up-type quarks in the box are different, there is no corresponding
L4 contribution (actually, there is, but that comes from four different 
$\lambda'$ type couplings which we neglect systematically), and the
treatment becomes simpler.

A note of caution here. In the literature, most of the bounds on the RPV
couplings are not absolute, but scale with the corresponding sfermion mass
that comes in the propagator. For a product coupling, the dependence is
typically $\lambda'\lambda'/m_{\tilde f}^2$. The results are usually
quoted for $m_{\tilde f} = 100$ GeV, and bounds for higher $m_{\tilde f}$
are straightforward to obtain. However, for the L4 boxes, there can be 
two distinct type of amplitudes, one through a quark-slepton box, and the other
through a squark-lepton box. For our analysis, we take all sleptons 
to be degenerate at 100 GeV, and all squarks to be degenerate at 300 GeV 
(100 GeV squarks are already ruled out). Thus, it is not easy to scale the
bounds quickly; a good way is to remember (and scale accordingly) 
that at the amplitude level, the
squark contribution is about 10\% of the slepton contribution, and the
second way, to get a more rough scaling, is to neglect the squark contribution
altogether.  
\begin{figure*}
\begin{center}
\centerline{\hspace*{3em}
\epsfxsize=14cm\epsfysize=4.5cm
                     \epsfbox{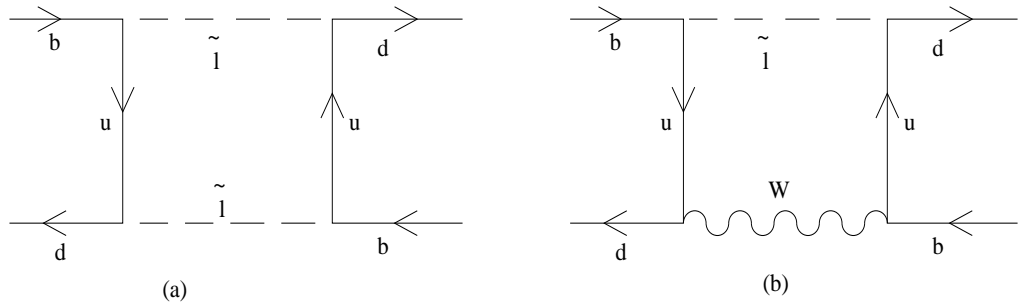}
}
\end{center}
\hspace*{-3cm}
  \caption{\em R-parity violating contributions to $\bbbar$ mixing.
Figure (a) corresponds to L4, while figure (b) to L2 amplitudes (see text for
their meanings). For L4, there are similar diagrams with squarks and leptons 
(both charged and neutral), as well as diagrams with left-chiral quarks
as external legs and quarks and sneutrinos flowing in the box.
For L2, there are diagrams where
the $W$ is replaced by the charged Higgs or the charged Goldstone. The internal
slepton can be of any generation, and so can be the internal charge $+2/3$
quarks, generically depicted as $u$.}
\end{figure*}

The effective Hamiltonian for the L4 box has been computed in 
\cite{decarlos-white}, with degenerate sfermions. For a nondegenerate case,
the expression reads ($q=2$ for $\kkbar$ and $q=3$ for $\bbbar$ boxes)
\begin{eqnarray}
{\cal H}_{L4} &=& {({\l'_{ik1}}^*\l'_{ikq})^2\over 128 \pi^2}\times\nonumber\\
&{}&\left[{1\over{\tilde m_l}^2} \left\{ I\left({m_{q_k^u}^2\over 
{\tilde m_l}^2}\right)
+ I\left({m_{q_k^d}^2\over {\tilde m_l}^2}\right)\right\}
+2{1\over{\tilde m_q}^2}\right]\tilde{O_1}\nonumber\\
&{}&\end{eqnarray}
where 
\begin{equation}
I(x) = {1-x^2+2x\log x\over (1-x)^3}
\end{equation}
with $I(0)=1$, which justifies the omission of this function in the last
term, where we have explicitly assumed all leptons to be massless. In fact,
this is also a reasonable assumption for all quarks except the top quark.

There may be another L4-type amplitude, with left-handed quarks as external legs
and right-handed down-type squarks and neutrinos (and quarks and sneutrinos
too) flowing inside the box. Neglecting the fermion masses, the effective
Hamiltonian reads
\begin{equation}
{\cal H}'_{L4} = {({\l'_{i1k}}^*\l'_{iqk})^2\over 128 \pi^2}
\left[ {1\over{\tilde m_l}^2} 
+{1\over{\tilde m_q}^2}\right]\tilde{O_1}.
\end{equation}
Note that this bounds a different combination. As we will see soon, there
is a corresponding L2 box with the same coupling as in ${\cal H}_{L4}$,
but none for ${\cal H}'_{L4}$. (The reason is simple: $W$-bosons couple only
to left-chiral quarks, and scalar couplings are negligible.) Thus, all
$\l'_{i1k}\l'_{iqk}$ combinations are expected to yield the same bound.

The case for the L2 box is more complicated. A simplified treatment, keeping
all bosons (sleptons, Ws, and Higgs bosons) to be degenerate at 100 GeV,
neglecting the QCD corrections, and saturating the experimental number with
RPV amplitude alone, was given in \cite{gg-arc}. Only real RPV couplings were
considered since there was no data for CP violation in the B system at that
time. We improve the calculation by incorporating all the above factors, but
note that the bounds quoted in \cite{gg-arc} for $\tan\beta=1$ are still
surprisingly close to the mark. In fact, with the introduction of
RPV phases, the possibility of destructive interference between the amplitudes
opens up, and one would have expected a weaker bound than in \cite{gg-arc}.
Consideration of the CP violating parameters $\sin 2\beta$ and $\varepsilon_K$
helps us to put a tighter set of constraints. 
The effective Hamiltonian, in this approximation,
is given by
\begin{eqnarray}
{\cal H}_{L2} &=& -{G_F {\l'}_{ik1}^*{\l'}_{ikq}\over 4\sqrt{2}\pi^2} V_{kq}^*
V_{kd}\times \nonumber\\
&{}& \left[ (1+\cot^2\beta) x_k^2 J(x_k) + I(x_k)\right] O_4,
\end{eqnarray}
where $x_k = m_{u_k}^2/m_{\tilde l}^2$, and
\begin{equation}
J(x) = {-2(x-1)+(x+1)\log x\over (x-1)^3},
\end{equation}
and $\cot\beta = v_d/v_u$, the ratio of the vacuum expectation values of
the two Higgs bosons that give mass to the down- and the up-type quarks
respectively (not to be confused with the phase of $V_{td}$). For
nondegenerate masses the expression is expectedly more complicated:
\begin{eqnarray}
{\cal H}_{L2}&=& -{G_F {\l'}_{ik1}^*{\l'}_{ikq}\over 4\sqrt{2}\pi^2} V_{kq}^*
V_{kd}\times\nonumber\\
&{}&\left[{m_W^2}{\cal A}_{L2}^{W}+{\cot^2\beta}{m_k^4}{\cal A}_{L2}^{H}+
{m_k^4}{\cal A}_{L2}^{G}\right] O_4,\nonumber\\
&{}&
\end{eqnarray}
where ${\cal A}_{L2}^{W,H,G}$s are factors that come out of the box diagram 
integration (the superscripts indicate the SM boson in the loop). Note that
as expected, the Higgs- and Goldstone-box contributions are negligible
for $k=1,2$.
The expressions for these factors in 't Hooft-Feynman
gauge are given in Appendix A.

The Hamiltonian is slightly modified if we consider two different up-type
quarks $k$ and $p$ inside the loop:
\begin{eqnarray}
{\cal H}_{L2}&=& -{G_F {\l'}_{ik1}^*{\l'}_{ipq}\over 4\sqrt{2}\pi^2} V_{kq}^*
V_{pd}\times\nonumber\\
&{}&\left[{m_W^2}{\cal B}_{L2}^W+{\cot^2\beta}m_k^2 m_p^2
{\cal B}_{L2}^H+m_k^2m_p^2{\cal B}_{L2}^G\right]O_4,\nonumber\\
&{}&
\end{eqnarray}
and again the integration factors ${\cal B}_{L2}^{W,G,H}$ are shown explicitly 
in Appendix A. Note that the Hamiltonian is evaluated at $q^2=m_W^2$ and
the matrix elements are to be evaluated at the proper low-energy scale.

Apart from these new amplitudes, there are a number of other boxes, a detailed
list of which is given in Appendix B, which are proportional to four
$\l'$ type couplings. They may be important in a scheme with a definite texture
for such couplings, but for our present study, we do not consider them 
any further.

\section{Numerical Inputs}

The major sources of the numerical inputs are: (i) the Heavy Flavor
Averaging Group (HFAG) website \cite{hfag} for the latest (summer 2003)
updates on B physics; (ii) Particle Data Group 2002 edition \cite{pdg2002}
and update for 2004 available on the web \cite{pdg2003}; and
(iii) the inputs used in the CKMfitter package \cite{ckmfitter}. The
quark masses and Wilson coefficients have been taken from \cite{becirevic,
ciuchini}. We use the following numbers.

Masses (all in GeV) \cite{becirevic,ciuchini,ckmfitter}:
\begin{eqnarray}
&{}&m_B = 5.2794; \ \ m_K = 0.494; \ \ m_b = 4.23;\nonumber\\ 
&{}&m_t = 167\pm 5;\ \ m_c = 1.3;\ \ m_s(2~{\rm GeV}) = 0.125;\nonumber\\
&{}&m_d = 0.007.
\end{eqnarray} 
The quark masses have been evaluated in the $\msbar$ scheme. The pole
mass for the top quark is about 5 GeV higher and the mass for the bottom
quark is $4.6$ GeV. The strange quark mass, however, is a major source of 
error for the chirally enhanced operators $O_4$ and $O_5$ evaluated for the 
$\kkbar$ system. The function $S_K$ increases by a factor of 2 if we 
take $m_s=95$ MeV, which in turn tightens the bound on the $\l'\l'$
couplings. $m_s=125$ MeV is almost at the uppermost limit of the allowed
range and hence generates the most conservative bound.   

The mass differences, in ps$^{-1}$, are given by \cite{hfag,pdg2002}
\begin{equation}
\Delta m_B = 0.502\pm 0.006;\ \ \Delta m_K = (5.31\pm 0.01)\times 10^{-3}.
\end{equation}

The CP-violating parameter $\sin(2\beta)$ is taken as the average over
all charmonium modes \cite{hfag}, since we have reasons to suspect that
there may be hints of new physics in $b\to s$ transition. On the other hand,
the theoretical prediction of $\sin(2\beta)$, obtained from a fit excluding
the direct experimental results, is taken from \cite{0307195}:
\begin{eqnarray}
\sin(2\beta)_{exp} &=& 0.736 \pm 0.049;\nonumber\\
\sin(2\beta)_{th} &=&  0.685 \pm 0.052.
\end{eqnarray}
We also use \cite{ckmfitter}
\begin{equation}
|\varepsilon_K| = (2.282\pm 0.017) \times 10^{-3}.
\end{equation}

This brings us to the only possible 
caveat of our analysis. If we assume the existence
of NP in the $\bbbar$ mixing amplitude, one should not use $\Delta m_B$
data for extracting $V_{td}$. In fact, what is needed is a reevaluation of the
bounds excluding all inputs that may be affected from NP ({\em e.g.},
$\Delta m_B$, $\epsilon_K$, etc.). We do not venture into that here. Rather,
we take the 90\% confidence limit (CL) bound on $V_{td}$ assuming unitarity
of the CKM matrix and considering tree-level processes only \cite{pdg2003}: 
\begin{equation}
|V_{td}| = 0.0094\pm 0.0046.
\end{equation}
Rest of the CKM elements are taken from \cite{pdg2003}, most of them
at their central value:
\begin{eqnarray}
&{}&|V_{ud}|=0.9745,|V_{us}|=0.224,|V_{ub}|=0.0037(8),
\nonumber\\
&{}&|V_{cd}|=0.224, |V_{cs}|=0.9737, |V_{cb}|=0.0415,\nonumber\\
&{}&|V_{ts}|=0.040(3), |V_{tb}|=0.99913.
\end{eqnarray}
The angle $\gamma$ is taken to lie between $50^\circ$ and $72^\circ$ 
\cite{ckmfitter}. The analysis is fairly insensitive to its precise value. 

The long- and short-distance QCD corrections to the box amplitudes are mostly
determined from lattice studies. We use \cite{ckmfitter}
\begin{eqnarray}
&{}& B_K = 0.86\pm 0.14 \pm 0.06,\nonumber\\
&{}& \sqrt{B_B} f_B = (0.228\pm 0.033)~{\rm GeV},\nonumber\\
&{}& f_K = 159.8~{\rm MeV},\nonumber\\
&{}& \eta_K = 1.38\pm 0.53;\ \ \eta_B = 0.55.
\end{eqnarray}

The leptonic and semileptonic BRs for the K meson, which are
of interest to us, are as follows \cite{pdg2003}:
\begin{eqnarray}
{\rm Br}(K^+\to \pi^+\nu\bar{\nu}) &=& 1.6^{+1.8}_{0.8}\times 10^{-10};
\nonumber\\
{\rm Br}(K^+\to \pi^+ \mu^+\mu^-) &=& (8.1\pm 1.4)\times 10^{-8};
\nonumber\\
{\rm Br}(K^+\to \pi^+ e^+ e^-) &=& (2.88\pm 0.13)\times 10^{-7};
\nonumber\\
{\rm Br}(K_L\to \mu^+\mu^-) &=& (7.24 \pm 0.14) \times 10^{-9}.
\end{eqnarray}
There are stringent bounds on $K_L\to e^+e^-$ too but that is chirally
suppressed compared to the $\mu^+\mu^-$ mode. For the B mesons, the relevant
numbers may be found in \cite{hfag}. We will not analyze them here (see
\cite{jyoti1} for such an analysis) but the bounds on RPV couplings 
obtained from B decays are not compatible with those obtained here, particularly
for a 300 GeV squark.

To evaluate the QCD corrections, we take $\alpha_s(m_Z^2) = 0.1172\pm
0.0020$ \cite{pdg2002}, and take the SUSY scale $M_S = 500$ GeV. The precise
value of this scale is not important, however, and we can take it to be at
the squark mass scale (300 GeV) without affecting the final results. The
exact evolution matrix can be found in \cite{ciuchini} and \cite{becirevic};
for our purpose, it is sufficient to note that for the K system, the operator
$\tilde{O_1}$ is multiplicatively renormalized by a factor 0.794 at the 
scale $\mu=2$ GeV, and the operator $O_4$ at $m_W$ changes to 
$(3.965 O_4 + 0.149 O_5)$. For the B system, the respective numbers are
$\tilde{O_1}\to 0.820 \tilde{O_1}$, $O_4\to (2.83 O_4 + 0.077 O_5)$. 
We again stress that theoretically the procedure is questionable
for boxes with light quarks flowing in the loop. However, the numbers that
we obtain are fairly robust and one can very well drop the NLO
corrections altogether, if necessary, without compromising the results.

The relevant $B$-parameters (eq. (\ref{bparam})) are \cite{lattice-b,
ciuchini}
\begin{eqnarray}
{\rm For~B}:&{}& \nonumber\\
&{}& B_1(m_b) = 0.87(4)^{+5}_{-4},\nonumber\\
&{}& B_4(m_b) = 1.16(3)^{+5}_{-7},\nonumber\\
&{}& B_5(m_b) = 1.91(4)^{+22}_{-7},\nonumber\\
{\rm For~K}:&{}& \nonumber\\
&{}& B_1(\mu) = 0.60(6),\nonumber\\
&{}& B_4(\mu) = 1.03(6),\nonumber\\
&{}& B_5(\mu) = 0.73(10).
\end{eqnarray}
Since the $O_5$ admixture is small, one can take the central values
for these parameters without introducing too much error.

In the supersymmetry sector, we assume only one RPV product coupling to be
nonzero at a time. We take all sleptons to be degenerate at 100 GeV, and all
squarks at 300 GeV. We also take $\tan\beta (\equiv v_2/v_1) = 5$ (very low
values are excluded by LEP, and the numbers are not sensitive to the precise
choice of $\tan\beta$), and the charged Higgs boson mass as 200 GeV (lower 
values are disfavored from $b\to s\gamma$).

\section{Analysis}
Our bounds are summarized in tables 1 and 2, for the K and the B
systems respectively. The procedure has been outlined in Section 2,
and is identical for both systems. However, for the $\kkbar$ system,
the SD contribution may not be the full story.
   If the SD contribution is smaller, the contribution to $\Delta m_K$ is also
   smaller, and hence the analysis runs for a smaller $\Delta m_K$ and gives
   tighter bounds on the relevant RPV couplings. The bounds we obtain are, 
therefore, the most lenient ones. 

It appears that the constraints on the $\kkbar$ system are essentially
controlled by $\varepsilon_K$, particularly when we allow for complex
RPV couplings. In the SM, the CP violating part is suppressed by the CKM
factors; with RPV there is no such suppression, and thus one gets tight
constraints. They are shown in the third and fourth columns of table 1. The
second column shows the constraints for real RPV couplings. Unless the L2
box contains a CKM phase, the imaginary part coming from RPV is zero, and
there is no bound coming from $\varepsilon_K$; the only constraint is
that coming from $\Delta m_K$. These entries have been marked with a dagger.

For the $\bbbar$ system,
one noteworthy thing is that the bounds on the real and the imaginary parts
of any product coupling are almost the same. (That is why we show only the
real part; the same bound applies to the imaginary part too. We found it to
differ by at most 10\% in a few cases.)
This is, of course, no numerical
accident. To understand this, let us analyze the origin of these bounds. 
There are two main constraints for the B system: $\Delta m_B$ and 
$\sin(2\beta)$. 
There will be a region, centered around the origin (since $\Delta m_B$ can be
explained by the SM alone) of $Re(\l'\l')-Im(\l'\l')$ plane, where $|\l'\l'|$
is small and the phase can be arbitrary. This is the SM-dominated region, 
where RPV creeps in to whatever place is left available. A conventional 
analysis, taking both SM and RPV but assuming incoherent sum of amplitudes,
should generate this region only.

\begin{table}
\begin{tabular}{||l|c|c|c|c||}
\hline
$\l'\l'$    & Only & Complex,  & Complex,   &Previous\\
combination & real & real part & imag. part &bound\\
\hline
(i31)(i32)$\spadesuit$ 
           & $4.5 \times 10^{-6}$ & $1.0 \times 10^{-5}$ & $5.0\times 10^{-6}$& 
0.203 \\
(i21)(i22) & $3.9\times 10^{-7}$ $\dag$
 & $4.0 \times 10^{-8}$ & $1.7\times 10^{-9}$& 
1.254 \\
(i11)(i12) & $3.9\times 10^{-7}$ $\dag$
 & $4.9 \times 10^{-8}$ & $1.6\times 10^{-9}$& 
0.109 \\
(i31)(i22) & $3.3\times 10^{-5}$ $\dag$
 & $5.5 \times 10^{-6}$ & $1.4\times 10^{-7}$& 
0.504 \\
(i21)(i32) & $5.1 \times 10^{-7}$ & $2.7 \times 10^{-6}$ & $1.1\times 10^{-6}$&
0.504 \\
(i21)(i12)$^\ddag$ & $9.0\times 10^{-8}$ $\dag$
              & $1.6 \times 10^{-8}$ & $4.0\times 10^{-10}$
& $10^{-9}$ \\
(i11)(i22) & $1.7\times 10^{-6}$ $\dag$
 & $4.9 \times 10^{-7}$ & $7.1\times 10^{-9}$&
 0.370 \\
(i31)(i12) & $7.5\times 10^{-6}$$\dag$
  & $2.6 \times 10^{-6}$ & $2.9\times 10^{-8}$&
 0.149 \\
(i11)(i32) & $2.2 \times 10^{-6}$ & $3.5 \times 10^{-5}$ & $1.1\times 10^{-5}$&
 0.149 \\
(i1j)(i2j)$\spadesuit$ 
           & $2.7\times 10^{-3}$ $\dag$
 & $2.3 \times 10^{-3}$ & $6.0\times 10^{-4}$&
 0.370 \\
\hline
\end{tabular}
\caption{Bounds on $\l'\l'$ combinations from $\kkbar$ mixing and
$\varepsilon_K$, shown for
cases when the product is strictly real and when the product may be complex.
The table displays the magnitudes only, and not the signs. 
Weakest DPBs, displayed in the last column, occur for 
$i=3$. They are from \cite{dreiner} except that we scale them for 
300 GeV squarks wherever necessary, unless they hit the perturbative bound. 
The entries marked with a dagger are bounded from $\Delta m_K$, see text.
The one marked with a double dagger is bounded from tree-level $\kkbar$
mixing, and those marked with a $\spadesuit$ have bounds of the order of
$10^{-5}$ from squark-mediated $K^+\to \pi^+\nu\bar{\nu}$ (see text).}
\end{table} 
\begin{table}
\begin{tabular}{||l|c|c|c||}
\hline
$\l'\l'$ & Only & Complex, & Previous\\
combination & real & real part & bound\\
\hline
(i31)(i33) & $1.7 \times 10^{-3}$ & $1.7 \times 10^{-3}$ & 0.203 \\
(i21)(i23) & $5.0 \times 10^{-5}$ & $1.6 \times 10^{-3}$ & 1.254 \\
(i11)(i13) & $8.0 \times 10^{-4}$ & $2.6 \times 10^{-3}$ & 0.109 \\
(i31)(i23) & $6.0 \times 10^{-6}$ & $1.6 \times 10^{-4}$ & 0.504 \\
(i21)(i33) & $0.13$ & $0.11$ & 0.504 \\
(i21)(i13) & $1.2 \times 10^{-5}$ & $2.9 \times 10^{-4}$ & 0.370 \\
(i11)(i23) & $2.6 \times 10^{-3}$ & $1.6 \times 10^{-2}$ & 0.370 \\
(i31)(i13) $^\dag$ & $2.8 \times 10^{-6}$ & 
              $2.5 \times 10^{-5}$ & $8\times 10^{-8}$ \\
(i11)(i33) & $2.7 \times 10^{-2}$ & $2.7 \times 10^{-2}$ & 0.149 \\
(i1j)(i3j) & $2.0 \times 10^{-3}$ & $4.1 \times 10^{-3}$ & 0.149 \\
\hline
\end{tabular}
\caption{Bounds on $\l'\l'$ combinations from $\bbbar$ mixing. Rest is same 
as table 1, except that the entry marked with a dagger
tree-level $\bbbar$ mixing.}
\end{table}

However, there is always scope for fully constructive or destructive 
interference. Consider a situation where the RPV contribution is large,
so large that even after a destructive interference with the SM amplitude,
enough is left to saturate $\Delta m_B$. This RPV-dominated region
(this is true for all NP models in general) gives us the bounds, and in the
limit where the SM can be neglected, the bounds on $Re(\l'\l')$ are almost 
the same as on $Im(\l'\l')$. 
The bounds are, however,
slightly different for the case where the RPV couplings are all real to the
case where they can be complex; the reason is the SM phase in mixing,
which chooses a particular direction in the $Re(\l'\l')-Im(\l'\l')$ plane
not coinciding with the real axis. This complication is absent for the
K system; the SM amplitude is almost real. The pattern can be understood from
Figures 2(a)-2(b).
It is obvious that the bound is controlled by the $\Delta m_B$ data. In 
Fig. 2(a) for the black boxes, the central cross is the region where 
L2 is important (and the interference with SM); however, the L4-dominated
region gives the bound. Fig. 2(b) shows a similar scan plot for the $\kkbar$
box driven by $\l'_{i31}\l'_{i32}$, but without $\varepsilon_K$ (this also
gives the reader an idea how strong the $\varepsilon_K$ constraint can be). 
However, this analysis holds even for the cases with only L2 boxes.

\begin{figure*}[htbp]
\vspace{-10pt}
\centerline{\hspace{-3.3mm}
\rotatebox{-90}{\epsfxsize=6cm\epsfbox{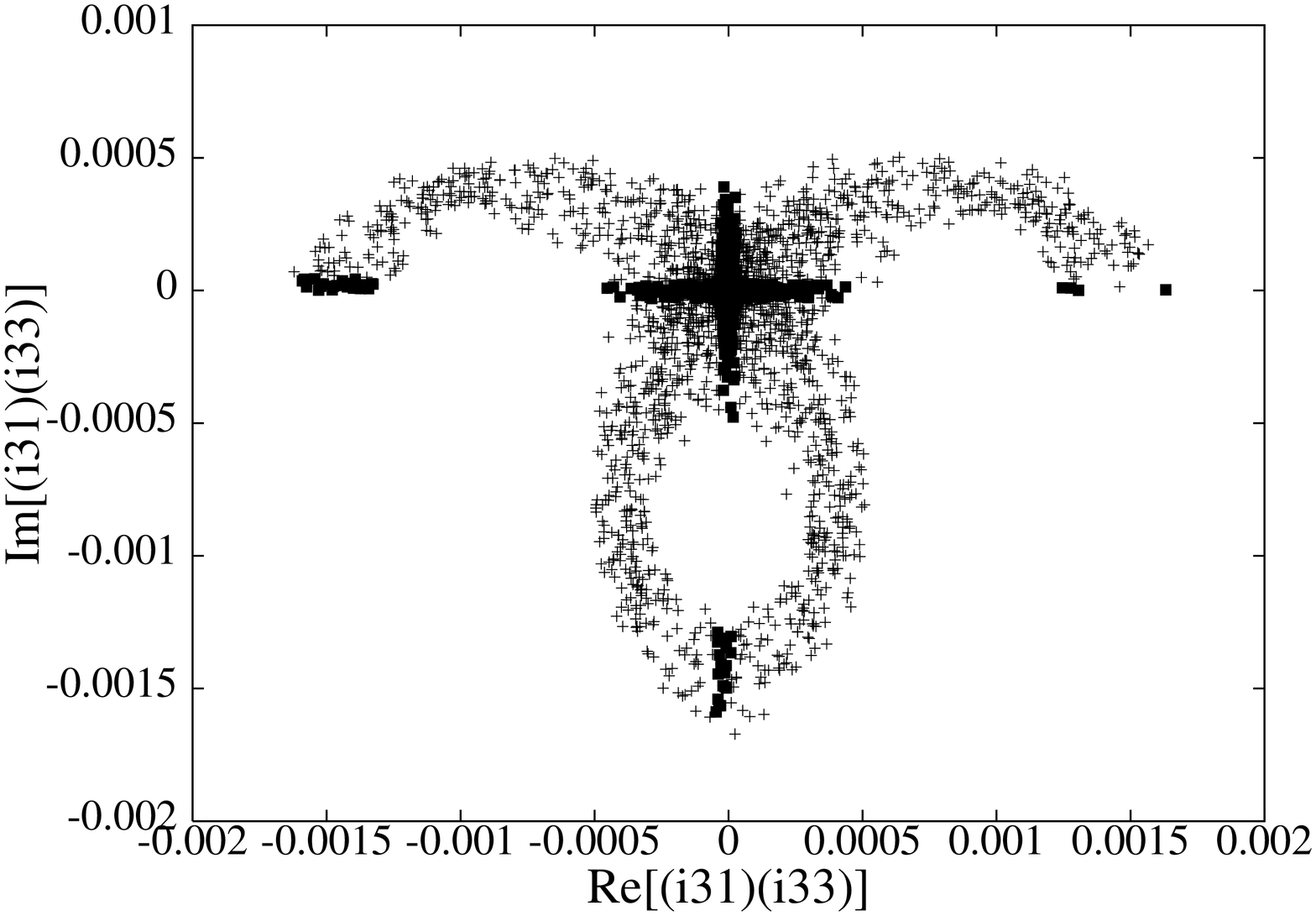}}
\hspace{-0.1cm}
\rotatebox{-90}{\epsfxsize=6cm\epsfbox{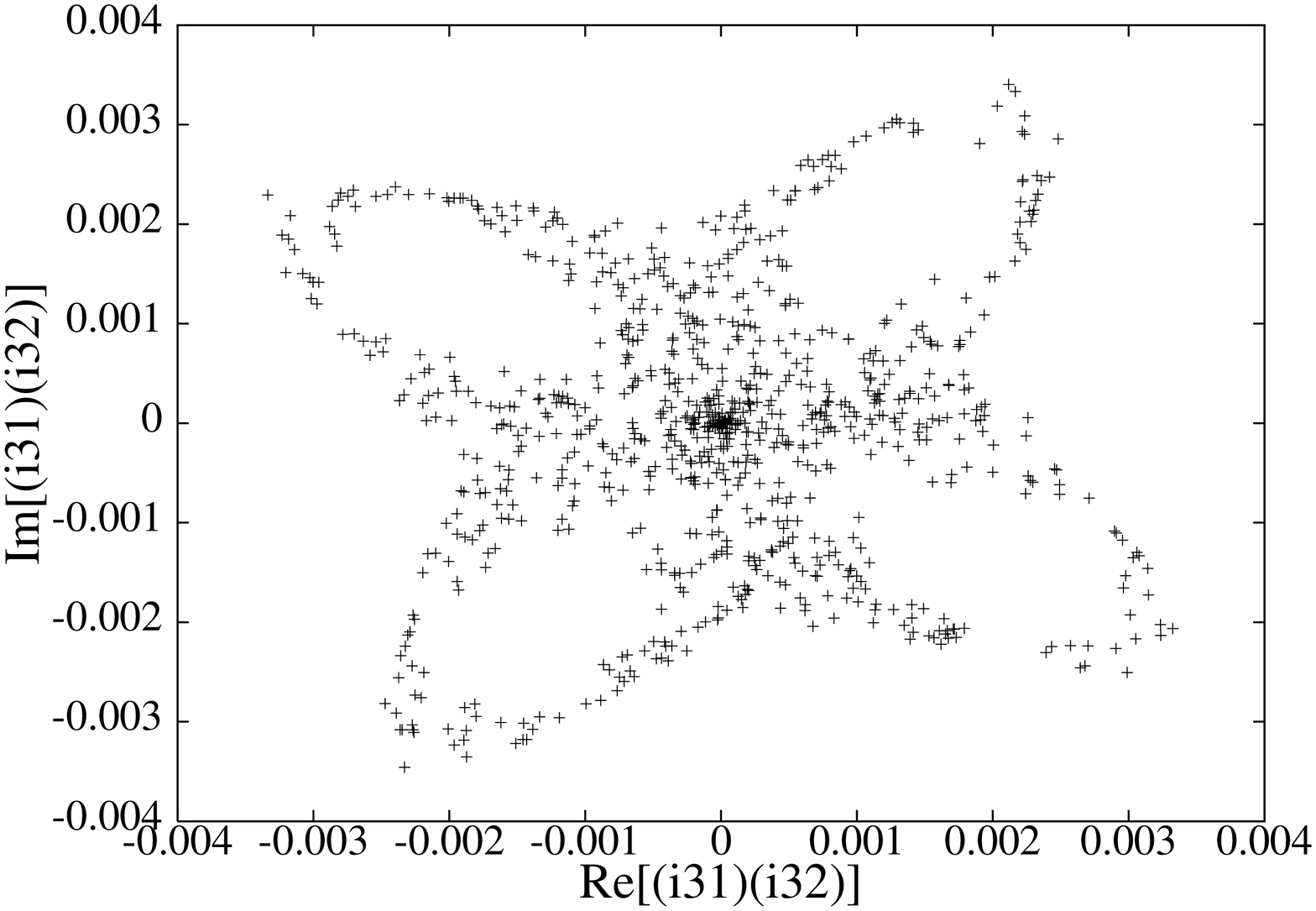}}}
\vspace*{3mm}
\centerline{\hspace{-0.5cm} (a) \hspace{7.5cm} (b)}
\hspace{3.3cm}
\caption{(a) Allowed parameter space for $\l'_{i31}\l'_{i33}$. The crosses
indicate the allowed parameter space switching the $\sin(2\beta)$ constraint 
off, and the black boxes do the same by keeping it, along with $\Delta m_B$.
(b) The parameter space for $\l'_{i31}\l'_{i32}$, which drives the $\kkbar$
mixing. Only $\Delta m_K$ constraint is used.}
\end{figure*}

The relative magnitude of the bounds is also easy to understand. Consider,
for example, the bounds on $\l'_{i31}\l'_{i23}$ vis-a-vis $\l'_{i21}\l'_{i33}$.
The relevant box diagrams have the same particle content; but the first one
is proportional to $V_{tb}V_{cd}$ ($\sim \lambda$), and the second one
to $V_{td}V_{cb}$ ($\sim {\cal O}(\lambda^5)$). The relative suppression 
in $\lambda$ enhances the limit on the RPV coupling.  

It is interesting to note that in most of the cases, the bounds are far
better than the DPBs. The exception is the product bounds responsible for
tree-level $\kkbar$ and $\bbbar$ mixing \cite{bsb}. 
Apart from them, $\l'_{i11}\l'_{i13}$ has a
comparable bound from $B^0\to\pi^+\pi^-$ \cite{bdk2}. 

Though mostly of the same orders of magnitude, these bounds are theoretically
an improvement over those obtained earlier \cite{decarlos-white,gg-arc,jyoti2}.
We have taken into account all possible amplitudes (and the interference
patterns play a nontrivial role), including the SM one, but have systematically
neglected the pure supersymmetric boxes coming from gaugino exchange. The
reason is that those boxes decouple in the heavy squark limit, and one can
always take an RPV model embedded in a minimal supersymmetric theory where 
such FCNC processes are somehow forbidden. It was shown in \cite{jyoti2} that
the bounds are fairly robust even if one takes into account such SUSY 
contributions.
Furthermore, the QCD corrections are implemented upto NLO. We have also
corrected a sign mistake (which, however, did not affect the result of the
earlier papers since there was no interference to be considered) in the
respective effective Hamiltonians. Lastly, we have also put bounds on
the imaginary parts of the couplings. As we have stressed earlier, the role
of the $\sin(2\beta)$ data is to banish a sizable portion of the
$Re(\l'\l')-Im(\l'\l')$ plane, rather than constraining the absolute 
bounds, see Fig. 2(a). For the K system, $\varepsilon_K$ chooses the region
about the origin. 

If we take a baryon-number violating RPV model, the only combinations
that can be constrained from the box amplitudes are $\l''_{i12}\l''_{i23}$
(from $\bbbar$ mixing) and $\l''_{i13}\l''_{i23}$ (from $\kkbar$ mixing).
The bounds are weaker by about two orders of magnitude from those on
their $\l'$ counterpart.

\section{Correlated Channels}

Though this paper is mainly on constraints on the $\l'\l'$ product couplings
coming from $\kkbar$ and $\bbbar$ boxes, let us also mention that such
couplings are also responsible for squark-mediated semileptonic ($b\to d\ell^+
\ell^-$, $s\to d\ell^+\ell^-$) and purely leptonic ($B^0\to\ell^+\ell^-$,
$K^0\to\ell^+\ell^-$) B and K decays (and slepton-mediated nonleptonic
decays) wherever kinematically possible. 

For the B mesons, no such leptonic mode has yet been observed. The 
corresponding upper limits on the BRs are of the order of
$10^{-7}$ for $\ell=e,\mu$ and $10^{-5}$ for the $\tau$ modes. With 300 GeV
squarks, from the bounds that one obtains here, a BR at most
of the order of $10^{-8}$ can be expected. Thus, we do not envisage to see
such leptonic channels before the next-generation hadronic or super $e^+e^-$
B factories. The semileptonic modes $B\to K^{(*)}\ell^+\ell^-$ have been
observed. However, the BRs are at the SM ballpark, and the RPV contributions
are expected to be smaller by at least one order of magnitude.

We have not discussed the $B_s$ system. That will evidently warrant a 
detailed analysis once the hadronic machines start running. Let us mention
that the channel $B_s\to\mu^+\mu^-$ can be mediated by a tree-level
squark exchange diagram, and with the present bound of $BR(B_s\to\mu^+\mu^-)
< 0.95\times 10^{-6}$ \cite{hfag}, the relevant upper limit on the 
RPV product coupling $\l'_{2i2}\l'_{2i3}$ is about $6\times 10^{-4}$
(for 300 GeV squarks). The limit obtained from the $\bsbsbar$ mixing should
be at the same level after a few years of data. Thus, one can have an 
interesting situation where there is a possibility of NP in both mixing and
decay.

The situation in the K meson system is better as far as the leptonic and
semileptonic modes are concerned. First, note that the four-Fermi effective
Hamiltonian for leptonic and semileptonic decays is of the form \cite{jyoti1}
\begin{eqnarray}
{\cal H}_{RPV} &=& {1\over 2} B_{jklm} \left[ \bar{\ell_j}\gamma^\mu P_L \ell_l
\right] \left[ \bar{d_m}\gamma_\mu P_R d_k\right] \nonumber\\
&{}& + {1\over 2} B_{jklm} \left[ \bar{\nu_j}\gamma^\mu P_L \nu_l
\right] \left[ \bar{d_m}\gamma_\mu P_R d_k\right] \nonumber\\
&{}& - {1\over 2} C_{jklm} \left[ \bar{\nu_j}\gamma^\mu P_L \nu_l
\right] \left[ \bar{d_k}\gamma_\mu P_L d_m\right] + H.c.,\nonumber\\
&{}&    \label{4fer}
\end{eqnarray}
where
\begin{equation}
B_{jklm} = \sum_{i=1}^3 { {\l'_{jik}}^*\l'_{lim}\over m_{\tilde 
{u/d}_{L_i}}^2 },
C_{jklm} = \sum_{i=1}^3 { {\l'_{jki}}^*\l'_{lmi}\over m_{\tilde d_{R_i}}^2 }.
\end{equation}
We take any one to be nonzero at a time. Assuming the leptonic phase-space
to be the same, and the $K^+\to \pi^+\nu\bar{\nu}$ to be dominated by RPV
SUSY, one obtains \cite{agashe}
\begin{equation}
{Br(K^+ \to \pi^+\nu\bar{\nu}) \over Br(K^+\to \pi^0 e^+\nu)}
= {|\l'_1\l'_2|^2 \over 4 G_F m_{\tilde d}^2 V_{us}^2}.
\end{equation}
Taking $Br(K_{e3}) = 0.0481$ (lower limit) and $Br(K^+\to \pi^+\nu\bar{\nu})
= 3.4\times 10^{-10}$ (1$\sigma$ upper limit), one obtains
\begin{equation}
|\l'_{ij1}\l'_{ij2}|, |\l'_{i1j}\l'_{i2j}| < 3.9\times 10^{-5}.
\end{equation}
It is clear that the modes $\pi 2e$ or $\pi 2 \mu$ will yield a less
severe bound, since the upper limit is weaker by orders of magnitude.

The decay $K_L\to \mu^+\mu^-$ bounds the imaginary part of the product
$\l'_{2i1}\l'_{2i2}$. The reason is that in the limit of CP conservation, 
$K_L = (K^0 - \bar{K^0})/\sqrt{2}$, and from eq. (\ref{4fer}), it is easy
to see that the real parts cancel out if we take the Hermitian 
conjugate term into account. The BR is
\begin{eqnarray}
Br(K_L\to\mu^+\mu^-)&=&{Im(\l'_{2i1}\l'_{2i2})^2\over 64\pi}\times\nonumber\\
&{}& \left( {f_K m_\mu
\over m_{\tilde u_i}^2}\right)^2 \sqrt{1-4{m_\mu^2\over m_K^2}} m_K \tau_{K_L},
\nonumber\\
&{}&
\end{eqnarray}
where $\tau_{K_L}$ is the lifetime of $K_L$. Putting the values, the bound is
\begin{equation}
Im(\l'_{2i1}\l'_{2i2})  < 3.5\times 10^{-5}.
\end{equation}
This is, of course, an approximate bound, since the SM has not been
included, and vacuum insertion is questionable for a t-channel squark
exchange diagram. 

In the K system, the nonleptonic decays are riddled with theoretical 
uncertainties. That is why one can always evade the bounds obtained from,
say, $\varepsilon'/\varepsilon$ \cite{abel}. On the other hand, we do
not have enough nonleptonic data from the B factories to put comparable 
bounds on the couplings we have considered here. The only exception is
$\l'_{i11}\l'_{i13}$, which can be bounded from the BR and the CP-asymmetry 
data of $B\to\pi^+\pi^-$. Other affected modes are listed in Table 3.
\begin{table}
\begin{tabular}{||l|c|c|c||}
\hline
$\l'\l'$ & Quark & Meson & UT\\
combination & level & level & angle\\
\hline
(i21)(i23) & $b\to c\bar{c} d$ & $B \to D^+D^-$ & $\beta$ \\
           & $b\to s\bar{s} d$ & $B \to K\bar{K}$ & $\beta$ \\
(i11)(i13) & $b\to u\bar{u} d$ & $B \to \pi(\rho) \pi(\rho)$ & $\alpha$ \\
(i21)(i13) & $b\to u\bar{c} d$ & $B \to D\pi$ & $\gamma$ \\
           & $b\to d\bar{s} d$ & $B \to \pi\bar{K}$ & SM suppressed \\
(i11)(i23) & $b\to c\bar{u} d$ & $B \to \bar{D}\pi$ & $\gamma$ \\
           & $b\to s\bar{d} d$ & $B \to K_S \pi$  & $\beta$ \\
(i1j)(i3j) & $b\to d\bar{s} s$ & $B \to \phi\pi$ &       \\
\hline
\end{tabular}
\caption{Some of the possible nonleptonic transitions mediated by
the RPV couplings discussed in the paper. We show only the more interesting
quark-level decays in column 2, and some typical decays in a generic manner
in column 3. The last column shows the UT angle that may be affected due
to the presence of RPV.}
\end{table}
However, we note that unless a product coupling is at least of the order of
$10^{-3}$, the RPV contribution is unlikely to affect the SM amplitude.
Still, a systematic study of NP is worthwhile; for example, the CP-asymmetry
in $B\to K_S\pi^0$, which is supposed to yield $\sin(2\beta)$, gives a
smaller central value \cite{babar-prelim}, notwithstanding the fact that the
experimental errors are large:
\begin{equation}
\sin(2\beta)(B\to K_S\pi^0) = 0.48^{+0.38}_{-0.47}\pm 0.06. 
\end{equation}
Note that this channel is mediated by a coupling which has a bound
of the order of $10^{-2}$ only.

\section{Summary and Conclusions}

In this paper, we have computed the bounds on the product of two 
RPV couplings of the type $\l'\l'$, coming from $\kkbar$ and $\bbbar$ mixing,
as well as from the $\sin(2\beta)$ constraint. Though such a calculation
is not new, we have implemented several features in the analysis which
have not been taken into account in earlier studies. We have considered
the exact expression for the box amplitudes, and have taken all possible
amplitudes, including that of the SM, into consideration. The QCD corrections
to the amplitudes have been taken upto the NLO level. We have considered
the possibility that the RPV product couplings may be complex. The
analysis is done in the benchmark point $m_{H^+}=200$ GeV, $\tan\beta = 5$,
all sleptons degenerate at 100 GeV and all squarks degenerate at 300 GeV,
and neglecting the pure MSSM contribution to the box amplitudes (by possibly
applying to some underlying FCNC suppression principle, like alignment
of the squark mass matrices). 

It is to be observed that in some cases, our bounds are actually weaker 
than those obtained earlier by saturating the mass difference with RPV
alone. The reason is that destructive interference with the SM amplitude
plays a very crucial role in determining the bounds, particularly when the
phase of the RPV coupling is arbitrary. 
There is an intricate interplay among different amplitudes
as can be seen in Fig. 2(a). All in all, we consider these bounds to be the 
most conservative (or in other words most robust) ones.

Some of these bounds can, however, be bettered if we consider the leptonic
and semileptonic K decays. Of course, one can enhance the squark mass
to a limit where these bounds become weaker than those obtained 
from the box (the latter is not much affected by decoupling the squarks), but
such extremely massive squarks are not interesting probably even for the 
Large Hadron Collider (LHC). There is no such competitive bound for the
B system, and one must wait for the hadronic B machines, or the super
$e^+e^-$ B factories. However, some of these couplings may affect the
nonleptonic decay modes (which, being slepton mediated, cannot be suppressed
by decoupling the squarks) of B, most important of them being 
$B\to\pi^+\pi^-$, which has been dealt with earlier \cite{bdk2}. 

The $B_s$ system is a different proposition. With the first data on 
$\Delta M_s$, one should try to compute all the possible ways NP can affect
this. Furthermore, one should also try to see whether such contributions
to mixing also naturally lead to sizable (and comparable with the SM)
contributions in decays like $B_s\to J/\psi \phi$, $\phi\phi$, $\phi\eta'$,
$\mu^+\mu^-$ et cetera. In this respect RPV emerges as an excellent
prospective candidate.

\begin{acknowledgments}
A.K. has been supported by the BRNS grant 2000/37/10/BRNS of DAE, Govt.\ of
India, and by the grant F.10-14/2001 (SR-I) of UGC, India. J.P.S. has been
supported by a SERC research fellowship.
\end{acknowledgments}
 
\appendix
\section{The kinematic functions coming from the box integrals}

The kinematic functions for the L2 boxes have been computed in the
't Hooft Feynman gauge, and we have explicitly checked that it is gauge
independent. 

First, let us deal the case where there is only one type of charge +2/3 
quark in the loop. We denote it by $q$. The bosons inside the box are a 
slepton (generically denoted as $\tilde l$) and one among the W boson,
the charged Higgs boson (of mass $m_h$) and the charged Goldstone boson 
(of mass $m_W$ in the 't Hooft-Feynman gauge). 

To start with, let us define some shorthand notations:
\begin{equation}
Q\equiv m_q^2,\ L\equiv m_{\tilde l}^2,\ W\equiv m_W^2,\ H\equiv m_h^2.
\end{equation}
The kinematic factors are
\begin{widetext}
\begin{eqnarray} 
{\cal A}^W_{L2} &=& {1\over (L - W) (L - Q)^2 (W - Q)^2} \times \nonumber\\ 
&{}&\Large[ Q (L - W) (2 LW - Q(L+W)) \log Q + L^2 (W - Q)^2 \log L\nonumber\\ 
&{}&+ (L - Q) \{ Q (W - Q) (L - W) - W^2 (L - Q) \log W\}\Large],\nonumber\\ 
{\cal A}^G_{L2} &=&
{\left[ W(L-Q)^2 \log W + (L-W) (Q^2-LW) \log Q
-(W-Q) \{ (L-W)(L-Q) + L (W-Q) \log L\}\right] \over
(L-W) (W-Q)^2 (L-Q)^2 },\nonumber\\
{\cal A}^H_{L2} &=& {\cal A}^G_{L2} (W\to H).
\end{eqnarray}  
The Hamiltonian is slightly modified if we have two different up-type
quarks $q_1$ and $q_2$. Using the shorthand $Q_1(Q_2)\equiv m_{q_{1,2}}^2$,
one has
\begin{eqnarray}
{\cal B}^W_{L2} &=& {1\over D}\left[ Q_1 F_1 \log Q_1 + Q_2 G_1 \log Q_2 +
L J_1 \log L -W K_1 \log W\right],\nonumber\\
{\cal B}^G_{L2} &=& -{1\over D}\left[ F_1 \log Q_1 + G_1 \log Q_2 +
J_1 \log L + K_1 \log W\right],\nonumber\\
{\cal B}^H_{L2} &=& {\cal B}^G_{L2} (W \to H).
\end{eqnarray}
Here
\begin{eqnarray}
D   &=& (Q_1-Q_2) (L-Q_1)(L-Q_2) (W-Q_1)(W-Q_2)(L-W),\nonumber\\
F_1 &=& Q_1 (L-Q_2) (W-Q_2) (L-W),\nonumber\\
G_1 &=& - F_1 (Q_1 \leftrightarrow Q_2),\nonumber\\
J_1 &=& (Q_1-Q_2) (W-Q_1) (W-Q_2) L,\nonumber\\
K_1 &=& - J_1 (L\leftrightarrow W).
\end{eqnarray}
\end{widetext}

We have checked that in the limit $Q_1=Q_2$, the ${\cal B}$ functions reduce
to corresponding ${\cal A}$s.
 
\section{The four-$\lambda'$ boxes}

Apart from the boxes that we have discussed, there may be boxes whose
amplitudes are proportional to the product of four different $\l'$
couplings. They are not important if we assume a hierarchical coupling
scheme where one and only one product is numerically significant, but may
appear in a scheme with a given texture at a high scale which results
in a number of nonzero couplings at the weak scale. Some of these couplings
have been shown in, {\em e.g.}, \cite{decarlos-white}.

The effective Hamiltonian is always proportional to, apart from the four
$\l'$ couplings, a factor of $1/64\pi^2$ times a momentum-space
integral $F(p^2)$ of the form
\begin{equation}
F(p^2) = \int{p^4\over \prod_{i=1}^4 (p^2+m_i^2)} dp^2
\end{equation}
where the denominator indicates the four propagators inside the box. To
obtain a quick numerical estimate, we may put all SM fermions except the
top quark to be massless, all sleptons (neutral and charged) to be degenerate
(say at 100 GeV) and all squarks to be degenerate too (say at 300 GeV).
The operator is either $O_1$ or $\tilde O_1$, see eq. (\ref{operators}).
If the sneutrinos are not their own antiparticles, there cannot be any
operators of the form $O_4$ or $O_5$.

In the following table, we show all possible combinations of the 4$\l'$
product, with the propagators. Expanded, they look like, {\em e.g.}, for
same up-type quarks but different sleptons: 
\begin{equation}
{\cal H}_{L4} = {\l'_{ijq}\l'_{kjq}{\l'}^*_{ij1}{\l'}^*_{kj1} \over 64\pi^2}
\tilde O_1 F(p^2)
\end{equation}
where $F(p^2)$ is to be evaluated with two quarks and two different sleptons.
\begin{table}
\begin{tabular}{||c|c|c||}
\hline
$4\l'$ & Internal & Operator\\
combination & propagators & \\
\hline
$(ijq)(kjq)(ij1)^*(kj1)^*$ & same $u$, different $\tilde\ell$  & $\tilde O_1$ \\
                           & same $d$, different $\tilde\nu$   & $\tilde O_1$ \\
                           & same $\tilde u$, different $\ell$ & $\tilde O_1$ \\
                           & same $\tilde d$, different $\nu$  & $\tilde O_1$ \\
$(i1j)(k1j)(iqj)^*(kqj)^*$ & same $d$, different $\tilde\nu$   &        $O_1$ \\
                           & same $\tilde d$, different $\nu$  & $       O_1$ \\
$(ijq)(ikq)(ij1)^*(ik1)^*$ & same $\ell$, different $\tilde u$ & $\tilde O_1$ \\
                           & same $\nu$, different $\tilde d$  & $\tilde O_1$ \\
                           & same $\tilde\ell$, different $u$  & $\tilde O_1$ \\
                           & same $\tilde\nu$, different $d$   & $\tilde O_1$ \\
$(i1j)(i1k)(iqj)^*(iqk)^*$ & same $\nu$, different $\tilde d$  & $       O_1$ \\
                           & same $\tilde\nu$, different $d$   & $       O_1$ \\
$(ijq)(lkq)(ik1)^*(lj1)^*$ & different $u$ and $\tilde\ell$    & $\tilde O_1$ \\
                           & different $d$ and $\tilde\nu $    & $\tilde O_1$ \\
                           & different $\ell$ and $\tilde u $  & $\tilde O_1$ \\
                           & different $\nu$ and $\tilde d $   & $\tilde O_1$ \\
$(i1k)(l1j)(iqj)^*(lqk)^*$ & different $d$ and $\tilde\nu $    & $       O_1$ \\
                           & different $\nu$ and $\tilde d $   & $       O_1$ \\
\hline
\end{tabular}
\caption{Different $4\l'$ combinations, and the corresponding 
propagators.}
\end{table}

Note that if these combinations are present, bounds coming from L2 boxes alone
have to be reevaluated.

\end{document}